\documentclass{ws-procs9x6}

\begin{document}

\title{Beam-size effect in Bremsstrahlung\footnote{\uppercase{T}his work is partially 
supported by grant 03-02-16154 of the \uppercase{R}ussian 
\uppercase{F}und of \uppercase{F}undamental \uppercase{R}esearch.}}

\author{V.~N. BAIER and V.~M.KATKOV}

\address{Budker Institute of Nuclear Physics, \\
Novosibirsk, 630090, Russia\\ 
E-mail: baier@inp.nsk.su; katkov@inp.nsk.su}

\maketitle

\abstracts{The smallness of the transverse dimensions of the colliding 
beams leads to suppression of bremsstrahlung cross section.
This beam-size effect was discovered and investigated at INP,
Novosibirsk. Different mechanisms of radiation are discussed. 
Separation of coherent and incoherent radiation is analyzed in detail.
For linear collider this suppression affects the whole spectrum.
It is shown that objections to the subtraction procedure in \cite{BK2} 
are groundless.}

\section{Introduction}

The bremsstrahlung process at high-energy involves a very small
momentum transfer. In the space-time picture this means that the
process occurs over a rather large (macroscopic) distance. The
corresponding longitudinal length (with respect to the direction
of the initial momentum) is known as the {\it coherence}
(formation) length $l_f$. For the emission of a photon with the
energy $\omega$ the coherence length is $l_f(\omega) \sim
\varepsilon(\varepsilon-\omega)/m^2\omega$, where $\varepsilon$
and $m$ is the energy and the mass of the emitting particle ( here
the system $\hbar=c=1$ is used). 

A different situation arises in the bremsstrahlung process at the
electron-electron(positron) collision. For
the recoil particle the effect turns out to be enhanced by the
factor $\varepsilon^2/m^2$. This is due to the fact that the main
contribution to the bremsstrahlung cross section gives the
emission of virtual photon with a very low energy 
$q_0 \sim m^2\omega/(\varepsilon (\varepsilon-\omega)) $ by the
recoil particle, 
so that the formation length of virtual photon is
$L_v(\omega)=l_f(q_0)= 4\varepsilon^3
(\varepsilon-\omega)/m^4\omega$.
This means that the effect for the recoil particles appears much
earlier than for the radiating particles and  can be important for
contemporary colliding beam facilities at a GeV range \cite{BK1}.

The  special experimental study of  bremsstrahlung was performed
at the electron-positron colliding beam facility VEPP-4 of
Institute of Nuclear Physics, Novosibirsk \cite{exp1}. The
deviation of the bremsstrahlung spectrum from the standard QED
spectrum was observed at the electron energy
$\varepsilon=1.84~$GeV. The effect was attributed to the smallness
of the transverse size of the colliding beams. In theory the
problem was investigated in \cite{BKS1}, where the bremsstrahlung
spectrum at the collision of electron-electron(positron) beams
with the small transverse size was calculated to within the power
accuracy (the neglected terms are of the order
$1/\gamma=m/\varepsilon$). After the problem was
analyzed in \cite{BD}, and later on in \cite{KPS} where some of 
results for the bremsstrahlung found in \cite{BKS1} were
reproduced.

It should be noted that in \cite{BKS1} (as well as in all other
papers mentioned above) an incomplete expression for the
bremsstrahlung spectrum was calculated. One has to perform the
subtraction associated with the extraction of pure fluctuation
process. This item will be discussed in Sec.2. In Sec.3 an 
analysis is given of incoherent radiation in electron-positron 
linear collider.

\section{Mechanisms of Radiation}

\subsection{Dispersion of momentum transfer}

We consider the radiation at head-on collision of high energy
electron and positron beams. The properties of photon emission
process from a particle
are immediately connected with details of its motion. It is
convenient to consider the motion and radiation from particles of
one beam in the rest frame of other beam (the target beam).
In this case the target beam is an ensemble of the Coulomb centers.
The radiation takes place at scattering of a particle from these
centers. If the target consists of neutral particles forming an
amorphous medium, a velocity of particle changes (in a random way)
only at small impact distances because of screening. In the
radiation theory just the random collisions are the mechanism
which leads to the incoherent radiation. For colliding beams
significant contributions into radiation give the large impact
parameters (very small momentum transfers) due to the long-range
character of the Coulomb forces. As a result, in the interaction
volume, which is determined also by the radiation formation length
(in the longitudinal direction), it may be large number of target
particles. Let us note that in the case when the contribution
into the radiation give impact parameters comparable with the
transverse size of target, the number of particles in the
interaction volume is determined by the ratio of the radiation 
formation length to the mean longitudinal distance between particles.

However, not all cases of momentum transfer should be interpreted
as a result of random collisions. One have to exclude the
collisions, which are macroscopic certain events. For elaboration
of such exclusion we present the exact microscopic momentum
transfer to the target particle in the form: ${\bf q}=<{\bf
q}>+{\bf q}_s$. Here $<{\bf q}>$ is the mean value of momentum
transfer calculated according to standard macroscopic
electrodynamics rules with averaging over domains containing many
particles. The longitudinal size of these domains should be large
with respect to longitudinal distances between target particles
and simultaneously small with respect to the radiation formation
length. The motion of particle in the averaged potential of
target beam, which corresponds to the momentum transfer $<{\bf
q}>$, determines the coherent radiation. While the term ${\bf
q}_s$ describes the random collisions which define the process of
incoherent radiation (bremsstrahlung). Such random collisions we
will call ``scattering'' since $<{\bf q}_s>=0$.

We consider for simplicity the case when the target beam is narrow, 
i.e.when the parameter $1/q_{min}$ characterizing the screening of
Coulomb potential in bremsstrahlung is much larger than the transverse
dimensions determining the geometry of problem. When a particle crosses the 
mentioned domain the transverse momentum transfer to particle is 
\begin{equation}
\textbf{q}= \sum_{a=1}^{N}\textbf{q}_a,\quad  
\textbf{q}_a=2\alpha
\frac{\mbox{\boldmath$\varrho$}-\mbox{\boldmath$\varrho$}_a}
{(\mbox{\boldmath$\varrho$}-\mbox{\boldmath$\varrho$}_a)^2},
\label{0.1}
\end{equation} 
where $\alpha$=1/137, $N$ is the number of particles in the 
domain under consideration of the counter-beam, 
$\mbox{\boldmath$\varrho$}$ is the impact parameter of particle, 
$\mbox{\boldmath$\varrho$}_a$ is the 
transverse coordinate of Coulomb center.
The mean value of momentum transfer is
\begin{equation}
\left\langle \textbf{q} \right\rangle = 
N\left\langle a|\textbf{q}_a |a\right\rangle,
\quad  \left\langle a|\textbf{q}_a|a \right\rangle =
\int \textbf{q}_a w_c(\mbox{\boldmath$\varrho$}_a)d^2\varrho_a, 
\label{0.2}
\end{equation} 
here $w_c(\mbox{\boldmath$\varrho$}_a)$ is the probability density of
target particle distribution over the transverse coordinates normalized 
to unity. Then
\begin{equation}
\left\langle \textbf{q}_s^2 \right\rangle =
N\left\langle a|\textbf{q}_a^2 |a\right\rangle +
N(N-1)\left\langle a|\textbf{q}_a |a\right\rangle^2-
\left\langle \textbf{q} \right\rangle^2
=N(\left\langle a|\textbf{q}_a^2 |a\right\rangle -
\left\langle a|\textbf{q}_a |a\right\rangle^2). 
\label{0.3}
\end{equation} 

The same expression for the dispersion of transverse momentum gives
the quantum analysis of {\it inelastic} scattering of emitting particle
on {\it a separate} particle of target beam 
in the domain under consideration 
\begin{eqnarray}
&& \overline{\textbf{q}_{in}^2}  =
\sum_{n \neq a}|\left\langle n|\textbf{q}_a |a\right\rangle|^2= 
\sum_{n}\left\langle a|\textbf{q}_a |n\right\rangle
\left\langle n|\textbf{q}_a |a\right\rangle
-\left\langle a|\textbf{q}_a |a\right\rangle^2
\nonumber \\
&&=\left\langle a|\textbf{q}_a^2 |a\right\rangle -
\left\langle a|\textbf{q}_a |a\right\rangle^2. 
\label{0.4}
\end{eqnarray} 

Granting that $N=n_z\Delta z$ we find for the mean value of 
transverse electric field of target beam
\begin{equation}
\textbf{E}_{\perp}(\mbox{\boldmath$\varrho$}, z)=
2 e n_z(z) \int \frac{\mbox{\boldmath$\varrho$}-\textbf{x}}
{(\mbox{\boldmath$\varrho$}-\textbf{x})^2} w_c(\textbf{x})d^2x 
\label{0.5}
\end{equation} 
For the rate of variation of function 
$\left\langle \textbf{q}_s^2 \right\rangle$ we get
\begin{equation}
\frac{d\left\langle \textbf{q}_s^2 \right\rangle}{dz}=
4\alpha^2 n_z(z)\left[\int \frac{w_c(\textbf{x})d^2x}
{(\mbox{\boldmath$\varrho$}-\textbf{x})^2}-
\left(\int \frac{\mbox{\boldmath$\varrho$}-\textbf{x}}
{(\mbox{\boldmath$\varrho$}-\textbf{x})^2}w_c(\textbf{x})d^2x \right)^2 \right]. 
\label{0.6}
\end{equation} 
We don't discuss here applicability of Eq.(\ref{0.6}) at 
$|\mbox{\boldmath$\varrho$}-\textbf{x}| \leq \lambda_c$
(see e.g. Sec.III in \cite{BK2})

It should be noted that in the kinetic equation which describes 
the motion of emitting particle
\begin{equation}
\frac{\partial f}{\partial t}+
\textbf{v} \frac{\partial f}{\partial \textbf{r}}+
\dot{\textbf{p}}\frac{\partial f}{\partial \textbf{p}}=
\mathbf{St} f
\label{0.7}
\end{equation} 
the value of electric field $\textbf{E}_{\perp}$ (\ref{0.5})
determines the coefficient $\dot{\textbf{p}}$ in l.h.s of equation
(\ref{0.7}) while r.h.s of equation arises due to random collisions
and is determined by Eq.(\ref{0.6}). The kinetic equation for 
description of radiation was first employed in \cite{M} and later
in \cite{BKS2} and \cite{BK3}. We consider the
case when the mean square angle of multiple scattering during the
whole time of beams collision is smaller than the square of
characteristic radiation angle. It appears, that this property is
sufficient for applicability of perturbation theory to the
calculation of the bremsstrahlung probability.

\subsection{Main characteristics of particle motion and radiation}

One of principal characteristics of particle motion defining the
properties of coherent radiation ({\it the beamstrahlung}) is the ratio
of variation of its transverse momentum to the mass during the
whole time of passage across the opposite beam $T$
\begin{equation}
\frac{\Delta {\bf p}_{\perp}}{m} \sim \frac{eE_{\perp}\sigma_z}{m}
\sim \frac{2\alpha N_c \lambda_c}{\sigma_x+\sigma_y} \equiv
\delta,
\label{2.1}
\end{equation}
where $N_c$ is the number of particles in the opposite beam,
$\sigma_x$ and $\sigma_y$ is its transverse dimensions ($\sigma_y
\leq \sigma_x$), $\sigma_z$ is the longitudinal size of opposite beam. 
The dispersion of particle momentum during time
$T$ is small comparing with $m$. It attains the maximum for the
coaxial beams:
\begin{equation}
\frac{\left<\textbf{q}_s^2\right>}{m^2} = \gamma^2
\left<\vartheta_s^2\right> \sim
\frac{8\alpha^2N_c\lambda_c^2}{\sigma_x\sigma_y}L \ll 1,
\label{2.2}
\end{equation}
here $\vartheta_s^2$ is the square of mean angle of multiple
scattering, $L$ is the characteristic logarithm of scattering
problem ($L \sim 10$). This inequality permits one to use the
perturbation theory for consideration of bremsstrahlung, and to
analyze the beamstrahlung independently from the 
bremsstrahlung\footnote{Actually more soft condition should be fulfilled:
\[
\left<\textbf{q}_s^2(l_f)\right>/m^2=\left<\textbf{q}_s^2\right>/m^2~l_f/\sigma_z \ll1
\]}.

Another important characteristics of motion is the relative
variation of particle impact parameter during time $T$
\begin{equation}
\frac{\Delta\varrho_i}{\varrho_i} \sim
\frac{eE_{\perp}\sigma_z^2}{\varepsilon \sigma_i} \sim
\frac{2\alpha N_c \lambda_c \sigma_z}{\gamma(\sigma_x+\sigma_y)
\sigma_i} \equiv D_i,
\label{2.3}
\end{equation}
here  $i$ is
$x$ or $y$. When the disruption parameter $D_i \ll 1$, the
collision doesn't change the beam configuration and the particle
crosses the opposite beam on the fixed impact parameter. If in
addition the parameter $\delta \ll 1$ (this situation is realized
in colliders with relatively low energies) then the beamstrahlung
process can be calculated using the dipole approximation. The main
contribution into the beamstrahlung give soft photons with an energy
\begin{equation}
\frac{\omega}{\varepsilon} \leq \frac{\gamma \lambda_c}{\sigma_z}
\ll 1.
\label{2.4}
\end{equation}

In the opposite case $\delta \gg 1$ the main part of
beamstrahlung is formed when the angle of deflection of particle
velocity is of the order of characteristic radiation angle
$1/\gamma$ and the radiation formation length $l_m$ is defined by
\begin{equation}
\frac{eE_{\perp}l_m}{m}\sim \frac{2\alpha N_c \lambda_c
l_m}{(\sigma_x+\sigma_y)\sigma_z}=1,\quad l_m =
\frac{\sigma_z}{\delta};
\label{2.5}
\end{equation}
and the characteristic photon energy is
\begin{equation}
\omega \sim \omega_m=\frac{\gamma^2}{l_m}
=\varepsilon \chi_m,\quad
\chi_m \equiv 2\alpha N_c
\gamma\frac{\lambda_c^2}{(\sigma_x+\sigma_y)\sigma_z} (\chi_m \ll
1).
\label{2.6}
\end{equation}
Here $\chi=|\mbox{\boldmath$\chi$}|$ is the invariant parameter
\begin{equation}
\mbox{\boldmath$\chi$}=\frac{\gamma}{E_0}
\left[{\bf E}_{\perp}+ {\bf v}\times{\bf H}\right],\quad 
{\bf E}_{\perp}={\bf E}-{\bf v}({\bf v}{\bf E}),
\label{2.7}
\end{equation}
where $E_0=m^2/e=1.32\cdot10^{16}$V/cm,
which defines properties of magnetic bremsstrahlung in the constant field
approximation (CFA). For applicability of CFA it is necessary that relative
variation of~ ${\bf E}_{\perp}$ in Eq.(\ref{0.5}) was small on the
radiation formation length $l_m$. As far $l_m$ is shorter
than $\sigma_z$ in $\delta \gg 1$ times  the characteristic parameter
becomes
\begin{equation}
D_{mi}=D_i\frac{l_m}{\sigma_z}=\frac{D_i}{\delta}=
\frac{\sigma_z}{\gamma \sigma_i},~ (i=x,y)
\label{2.8}
\end{equation}
to that extent.
The condition $D_{mi} \ll 1$ is fulfilled in all known cases. The
mean number of photons emitted by a particle during the whole
time of passage across the opposite beam $T$ is
$\left<N_{\gamma}\right> \sim \alpha\delta$, it include the
electromagnetic interaction constant. Using the estimate
(\ref{2.6}) we get an estimate of relative energy loss
\begin{equation}
\frac{\Delta\varepsilon}{\varepsilon} \sim \alpha\delta
\chi_m~(\chi_m \ll 1)
\label{2.9}
\end{equation}
In the case $\chi_m \ll 1$ (this condition is satisfied in all
existing facilities and proposed collider projects) the soft
photons with energy $\omega \sim \omega_m =\varepsilon\chi_m \ll
\varepsilon$ are mainly emitted. For $\omega \gg \omega_m$ the
emission probability is exponentially suppressed. So, such
photons are emitted in the bremsstrahlung  process only. The
boundary photon energy $\omega_b$, starting from which the
bremsstrahlung  process dominates, depends on particular parameters
of facility. If $\chi_m \sim 1/10$ the energy is $\omega_b \sim
\varepsilon$. The formation length for $\omega \gg \omega_m$ is
much shorter than $l_m$. On this length the particle deflection
angle is small comparing with $1/\gamma$ and one can neglect the
variation of transverse beam dimensions (see Eq.(\ref{2.8})).
This means that all calculations of bremsstrahlung characteristics 
can be carried out in adiabatic
approximation using local beam characteristics $\sigma_{x,y}(t),
{\bf v}(t)$ etc, with subsequent averaging of radiation
characteristics over time. Note that actually we performed a
covariant analysis and the characteristic parameters are defined
in a laboratory frame.

\subsection{Separation of coherent and incoherent radiation}

As an example we consider the situation when the configuration of
beams doesn't change during the beam collision (the disruption
parameter $D \ll 1$), and the total particle deflection angle
during intersection of whole beam is small comparing with the
characteristic radiation angle $1/\gamma$ (the dipole case). The
target beam in its rest frame is the ensemble of classical
potentials centers with coordinates ${\bf r}_a~({\bf x}_a,~z_a)$
and the transverse coordinate of emitting particle is ${\bf
r}_{\perp}$. In the perturbation theory the total matrix element
of the radiation process can be written as
\begin{equation}
\textbf{M}(\textbf{r}_{\perp})=\sum_{a=1}^{N_c}
\textbf{m} (\textbf{r}_{\perp}-\textbf{x}_a )\exp(iq_{\parallel} z_a)
\label{1.1}
\end{equation}
We represent the combination $M_iM_j^{\ast}$ in the form
\begin{eqnarray}
\hspace{-7mm}&& M_iM_j^{\ast}=\sum_{a=b}
m_i(\textbf{r}_{\perp}-\textbf{x}_a) m_j(\textbf{r}_{\perp}-\textbf{x}_b)
\nonumber \\
\hspace{-7mm}&& +\sum_{a \neq b}
m_i(\textbf{r}_{\perp}-\textbf{x}_a) m_j(\textbf{r}_{\perp}-\textbf{x}_b)
\exp(-iq_{\parallel}\left(z_a-z_b)\right).
\label{1.2}
\end{eqnarray}
In the expression Eq.(\ref{1.2}) we have to carry out averaging
over position of
scattering centers. We will proceed under assumption that there are
many scattering centers within the radiation formation
length $l_f=1/q_{\parallel}$
\begin{equation}
N_f=n_zl_f \gg 1,
\label{1.3}
\end{equation}
where for the Gaussian distribution
\begin{equation}
n_z =\frac{N_c}{\sqrt{2\pi}\sigma_z}
\exp\left(-\frac{z^2}{2\sigma_z^2} \right),
\label{1.4}
\end{equation}
here $N_c$ and $\sigma_z$ are introduced in Eq.(\ref{2.1}).
Note that in the situation under consideration $\varrho_{max}
=|\textbf{r}_{\perp}-\textbf{x}_a|_{max} \geq \sigma_t$, where
$ \sigma_t$ is the characteristic transverse size of target beam.
Let us select terms with approximately fixed phase
$q_{\parallel}(z_a-z_b)=\phi_{ab}$ in the sum with $a \neq b$ in Eq.(\ref{1.2}).
If the condition (\ref{1.3}) is fulfilled, there are many
terms for which the phase variation is small ($\Delta\phi_{ab} \ll 1$). For
this reason one can average over the transverse coordinates
(${\bf x}_a, {\bf x}_b$) of target particles in Eq.(\ref{1.2}) without
touching upon the longitudinal coordinates ($z_a, z_b$)
\begin{eqnarray}
\hspace{-7mm}&& M_iM_j^{\ast}=N_c \left\langle m_im_j\right\rangle_{\perp}
+\left\langle m_i\right\rangle_{\perp} \left\langle m_j\right\rangle_{\perp}
\sum_{a \neq b} \exp(-iq_{\parallel}\left(z_a-z_b)\right)
\nonumber \\
\hspace{-7mm}&&=N_c\Big( \left\langle m_im_j\right\rangle_{\perp}
-\left\langle m_i\right\rangle_{\perp} \left\langle m_j\right\rangle_{\perp} \Big)
+\left\langle m_i\right\rangle_{\perp} \left\langle m_j\right\rangle_{\perp}
\left|\sum_{a}\exp(iq_{\parallel}z_a)\right|^2,
\label{1.5}
\end{eqnarray}
where
\begin{eqnarray}
\hspace{-7mm}&&
\left\langle m_i\right\rangle_{\perp} =\int m_i(\textbf{r}_{\perp}-\textbf{x} )
w_c(\textbf{x})d^2x,
\nonumber \\
\hspace{-7mm}&&
\left\langle m_im_j\right\rangle_{\perp} =\int m_i(\textbf{r}_{\perp}-\textbf{x} )
m_j(\textbf{r}_{\perp}-\textbf{x} )w_c(\textbf{x})d^2x,
\label{1.6}
\end{eqnarray}
here $w_c(\textbf{x})$ is the probability density of target particle distribution
over the transverse coordinates normalized to unity. In Eq.(\ref{1.5}) in the
sum with $a\neq b$ we add and subtract the terms with $a=b$.
The first term (proportional to $N_c$) on the right-hand side of Eq.(\ref{1.5})
is the  incoherent contribution to  radiation ({\it the bremsstrahlung}).
The second term gives the coherent part of
radiation. For Gaussian
distribution  Eq.(\ref{1.4}) performing averaging over the longitudinal
coordinate $z_a$ one has
\begin{equation}
\left|\sum_{a}\exp(iq_{\parallel}z_a)\right|^2 \rightarrow
\left|\int_{-\infty}^{\infty} n_z\exp(iq_{\parallel}z)dz\right|^2=
N_c^2\exp(-q_{\parallel}^2\sigma_z^2).
\label{1.7}
\end{equation}

\section{Incoherent Radiation}

The correction to photon emission probability due to the small transverse
dimensions of colliding beam for unpolarized electrons and photon
was calculated in \cite{BK2} basing on
subtraction procedure as in Eq.(\ref{1.5}). It is obtained  
after integration over the azimuthal angle of the emitted photon 
\begin{equation}
dw_1=\frac{\alpha^3}{\pi
m^2}\frac{\varepsilon'}{\varepsilon}\frac{d
\omega}{\omega}U(\zeta) F(\omega, \zeta) d\zeta,\quad
\zeta = 1+\gamma^2 \vartheta^2, 
\label{3.11}
\end{equation}
where $\vartheta$ is the photon emission angle, 
$\varepsilon'=\varepsilon-\omega$,
\begin{eqnarray}
\hspace{-7mm}&& U(\zeta)=v-\frac{4(\zeta-1)}{\zeta^2},~ 
v=\frac{\varepsilon}{\varepsilon'}
 +\frac{\varepsilon'}{\varepsilon},\quad
F(\omega, \zeta)=F^{(1)}(\omega, \zeta)-F^{(2)}(\omega, \zeta),
\nonumber \\
\hspace{-7mm}&& F^{(1)}(\omega, \zeta)
=\frac{\eta^2}{\zeta^2}\int
\left[K_0(\eta \varrho)K_2(\eta \varrho)-K_1^2(\eta
\varrho)\right]\varrho
\frac{d\Phi(\mbox{\boldmath$\varrho$})}{d\varrho}d^2\varrho
\nonumber \\
\hspace{-7mm}&& F^{(2)}(\omega, \zeta)=\frac{2\eta^2}{\zeta^2}\int
\left(\int K_1(\eta
\varrho)\frac{\mbox{\boldmath$\varrho$}}{\varrho}w_c({\bf
x}-\mbox{\boldmath$\varrho$})d^2\varrho \right)^2w_r({\bf x})d^2x,
\label{3.12}
\end{eqnarray}
here
\begin{equation}
\eta=q_{min}\zeta,~ q_{min}=m^3\omega/4\varepsilon^2 \varepsilon',~
\Phi(\mbox{\boldmath$\varrho$})=\int w_r({\bf
x}+\mbox{\boldmath$\varrho$}) w_c({\bf x})d^2x,
\label{3.15}
\end{equation}
where $w_c({\bf x})$ is defined in (\ref{1.6}), $w_r({\bf x})$ is the same but
for the radiating beam,  value $q_{min}$ is defined in c.m.frame of colliding particles. 
The term $F^{(2)}(\omega, \zeta)$ is the
subtraction term. The total probability is $dw_{\gamma}=dw_0+dw_1$, 
where $dw_0$ is standard QED probability. The analysis in \cite{BK2} was
based on Eqs.(\ref{3.11})-(\ref{3.12}).

We considered in \cite{BK2} the actual case of the
Gaussian beams. The Fourier transform was used
\begin{eqnarray}
\hspace{-9mm}&&w({\bf x})=\frac{1}{(2\pi)^2}\int d^2q \exp(-i{\bf q x})w({\bf
q});
\nonumber \\
\hspace{-9mm}&& w_r({\bf
q})=\exp\left[-\frac{1}{2}(q_x^2\Delta_x^2+q_y^2\Delta_y^2)\right],~
w_c({\bf
q})=\exp\left[-\frac{1}{2}(q_x^2\sigma_x^2+q_y^2\sigma_y^2)\right],
\label{4.1}
\end{eqnarray}
where as above the index $r$ relates to the radiating beam and the
index $c$ relates to the target beam, $\Delta_y$ and $\Delta_x$
($\sigma_y$ and $\sigma_x$) are the vertical and horizontal
transverse dimensions of radiating (target) beam. Substituting
(\ref{4.1}) into Eq.(\ref{3.15}) we find
\begin{equation}
\Phi(\mbox{\boldmath$\varrho$})=
\frac{\Sigma_x\Sigma_y}{\pi}\exp[-\varrho_x^2\Sigma_x^2-\varrho_y^2\Sigma_y^2];~
\Sigma_x^2=\frac{1}{2(\sigma_x^2+\Delta_x^2)},~
\Sigma_y^2=\frac{1}{2(\sigma_y^2+\Delta_y^2)}. 
\label{4.2}
\end{equation}
Using the relation $d\sigma_1=\Phi^{-1}(0) dw_1$ the following expression
for the correction to spectrum was found in \cite{BK2} starting from (\ref{3.11})
\begin{eqnarray}
\hspace{-2mm}&& d\sigma_1^{(1)}=\frac{2\alpha^3}{m^2}
\frac{\varepsilon'}{\varepsilon}
\frac{d\omega}{\omega}f^{(1)}(\omega),~f(s) = \frac{\sqrt{\pi}}{2s}(v-8s^2){\rm
erfc}(s)+4\displaystyle{e^{-s^2}}+2{\rm Ei}(-s^2),
\nonumber \\
\hspace{-2mm}&& f^{(1)}(\omega)= -\frac{1}{\pi\Sigma_x
\Sigma_y}\int_{0}^{2\pi}\frac{d\varphi}{\Sigma_x^{-2}\cos^2\varphi+
\Sigma_y^{-2}\sin^2\varphi}
\int_{0}^{\infty}F_2(z)f(s)s ds,
\label{4.9} \\
\hspace{-2mm}&& z^2=\frac{s^2 q_{min}^{-2}}{\Sigma_x^{-2}\cos^2\varphi+
\Sigma_y^{-2}\sin^2\varphi},~ F_2(x)=\frac{2x^2+1}{x\sqrt{1+x^2}}
\ln(x+\sqrt{1+x^2})-1, 
\nonumber
\end{eqnarray}
where Ei($x$) is the exponential integral function and erfc($x$) is
the error function.
This formula is quite convenient for the numerical calculations.

The subtraction term ($F^{(2)}(\omega, \zeta)$ in (\ref{3.12})) gives
for coaxial beams
\begin{equation}
d\sigma_1^{(2)}=-\frac{2\alpha^3}{m^2}
\frac{\varepsilon'}{\varepsilon}
\frac{d\omega}{\omega}J^{(2)}(\omega), 
\label{4.19}
\end{equation}
where
\begin{eqnarray}
&& J^{(2)}(\omega)= \frac{\sqrt{a b}}{\Sigma_x
\Sigma_y}  \int_{0}^{\infty} ds_1 \int_{0}^{\infty} ds_2  g\left(
\frac{q_{min}\sqrt{s}}{2} \right)G(s_1, s_2),
\nonumber \\
&& G(s_1, s_2)=\frac{1}{2} \left( \frac{a_1a_2 b_1 b_2}{A
B}\right)^{1/2}\left[\frac{a_1a_2}{A}+\frac{b_1b_2}{B}\right]
\label{4.20}
\end{eqnarray}
Here the function $g(q)$ is:
\begin{equation}
g(q)= \left( v -\frac{2}{3}\right)\displaystyle{e^{-q^2}}-2q^2\left[
\frac{\sqrt{\pi}}{2q}\left(v-\frac{8}{3}q^2\right){\rm
erfc}(q)+\frac{4}{3}\displaystyle{e^{-q^2}}+{\rm Ei}(-q^2)\right].
\label{4.21}
\end{equation}
In (\ref{4.20}) we introduced the following notations
\begin{eqnarray}
&& a=\frac{1}{2\Delta_x^2}, \quad b=\frac{1}{2\Delta_y^2},\quad
 a_{1,2}=\frac{1}{s_{1,2}+2\sigma_x^2},\quad
b_{1,2}=\frac{1}{s_{1,2}+2\sigma_y^2},
\nonumber \\
&& A=a_1+a_2+a, \quad B=b_1+b_2+b,\quad s=s_1+s_2. 
\label{4.22}
\end{eqnarray}

In the case of narrow beams one has  $q_{min}/(\Sigma_x+\Sigma_y) \ll
1$. In this case of coaxial beams $d\sigma_{\gamma}= d\sigma_0 + d\sigma_1$ is
\begin{equation}
d\sigma_{\gamma}= 
\frac{2\alpha^3}{m^2} \frac{\varepsilon'}{\varepsilon}
\frac{d\omega}{\omega}\Bigg\{\left(v-\frac{2}{3}\right)\Bigg[2\ln
\frac{m}{\Sigma_x+\Sigma_y} +C +2
  -J_{-}\Bigg]+\frac{2}{9}\Bigg\},
 \label{5.7}
\end{equation}
where
\begin{equation}
 J_{-}=\frac{\sqrt{a b}}{\Sigma_x
\Sigma_y} \int_{0}^{\infty}
ds_1 \int_{0}^{\infty} ds_2 G(s_1, s_2)
\label{5.8}
\end{equation}

The dimensions of beams in the experiment \cite{exp1} were
$\sigma_y=\Delta_y=24~\mu m, \sigma_x=\Delta_x=450~\mu m$, so
this is the case of flat beams. The estimate for this case
gives $J_{-} \simeq (4/3\sqrt{3})\pi
\sigma_y/\sigma_x \ll 1$. This term is much smaller than other
terms in (\ref{5.7}). This means that for this case the
correction to the spectrum calculated in \cite{BKS1} is very
small. The parameters of beam in the
experiment \cite{P} were (in our notation): $\sigma_y=\Delta_y=(50 \div
58) \mu m$, $\sigma_x=\Delta_x=(250 \div 290) \mu m$. 
Since the ratio of the vertical and the horizontal dimensions is not 
very small, the contribution of subtraction term (Eq.(\ref{4.19})) is 
essential (more than 10\%).
For details of  comparison of experimental data \cite{exp1}, \cite{P} 
with theory see \cite{BK2}, where we discussed also possible
use of beam-size effect for linear collider tuning. It should be noted that
for linear collider the condition of strong beam-size effect 
$\sigma_y q_{min} \ll 1~(\sigma_y \ll \sigma_x)$ is fulfilled for  
the whole spectrum. This can be seen in Fig.1, where the lower curve 
is calculated using Eq.(\ref{4.9}) and the subtraction term is very
small since $\sigma_y/\sigma_x < 0.01$. As far as the narrow beams are
considered in Fig.1, the lower curve is consistent also 
with Eq.(\ref{5.7}). This curve depends on the energy and the
transverse sizes of beams. It will be instructive to remind that
the analysis in \cite{BK2} (see Eq.(2.8)) and here is valid if 
$\chi_m/u \ll 1$ (see Eq.(\ref{2.6}), $u=\omega/\varepsilon'$).
The parameter $\chi_m$ depends also on number of particles $N_c$
and the longitudinal beam size. So, for low $N_c$ Fig.1 is valid
for any $x$, but for TESLA project ($\chi_m$=0.13) it holds 
in hard part of spectrum only. In fact, the probability of
incoherent radiation becomes larger than the probability of
coherent radiation only at $x > 0.7$ where the lower curve in Fig.1
is certainly applicable.

\begin{figure}[ht]
%\epsfxsize=10cm   %width of figure - will enlarge/reduce the figures
%\epsfbox{fig3.eps}
%\figurebox{2cm}{3cm}{} %to have a box alone 
\centerline{\epsfxsize=3.9in\epsfbox{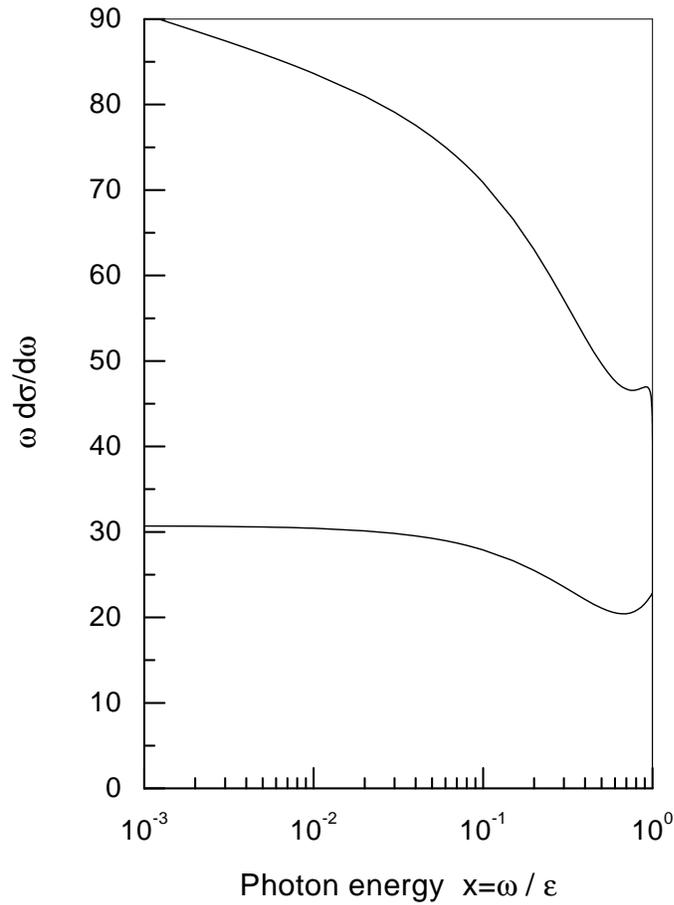}}   
\caption{The bremsstrahlung intensity spectrum $\omega
d\sigma/d\omega$ in units $2\alpha r_0^2$ versus the photon
energy in units of initial electron energy
($x=\omega/\varepsilon$) for linear collider with beam energy
$\varepsilon=250~$GeV. The upper curve
is the standard QED spectrum. The curve below is
calculated  with the beam-size effect taken into account for
$\sigma_x=553nm, \sigma_y=5nm$. \label{Fig.1}}
\end{figure}

Beam-size effect was considered using different approaches which are more
or less equivalent formulations of QED perturbation theory where the incident
particles consists of wave packets. In \cite{BKS1} the universal method was
used which permits to obtain any QED cross section within the relativistic 
accuracy (up to terms $\sim 1/\gamma$). A ``general scheme'' in \cite{KPS}
doesn't fall out the scope of \cite{BKS1} and particular derivation follows
method used in \cite{BD}. As it was shown above, the subtraction procedure
is necessary to extract pure fluctuation process, this was done in  
\cite{BK2}.

Recently in \cite{KS} this subtraction procedure 
was questioned.  An objection is based on correlator Eq.(28) in \cite{KS}. 
If one takes integrals over
${\bf r}$ and ${\bf r}'$ from both sides of this correlator, 
one obtains using Eq.(21) in \cite{KS}:$N_p^2=N_p(N_p+1)$. It is 
evident that the last relation
as well as correlator Eq.(28) in \cite{KS} are not adequate for the 
discussed problem since the subtraction term is of the same 
($\propto N_p$)as the relation error. 
According to \cite{KS} (see text before Eq.(19)) the correlator Eq.(28) is
obtained as result of ``the averaging over fluctuation of particle
in the field connected, for example, with the fluctuations of particle 
positions {\it for many collisions of bunches in a given experiment}''
(our italics BK). This statement has no respect to the problem under 
consideration. As it is shown in Secs.2.1 and 2.3 the main aspect is
{\it the presence of many scattering centers within the radiation
formation length}. An analysis in Appendix A confirms this conclusion
for radiation in crystals. The reference (Ref.21 in \cite{KS}) 
to the textbooks is 
senseless because different problems are discussed in these books.

The only correct remark in \cite{KS} 
is that in \cite{BK2} there was no derivation of the starting formulas. 
This derivation is given here above. In Sec.2.1 a generic picture of particle
motion in the field of counter-beam (in its rest frame) is given. 
A smooth variation
of transverse momentum in an averaged field of counter-beam is considered.
It determines the coherent radiation (the beamstrahlung).
 Along with smooth variation there are
the fluctuations of particle velocity due to multiple scattering on
the formation length. Just these fluctuations ensure the incoherent 
radiation (the bremsstrahlung). The mean square of transverse momentum
dispersion at multiple scattering on the formation length (see 
Eqs.(\ref{0.3}), (\ref{0.4}), (\ref{0.6})) determines the 
bremsstrahlung probability. The mentioned equations contain both
the singular and the subtraction terms accordingly to \cite{BK2}.
In Sec.2.3 we  separated the coherent and incoherent parts
of radiation explicitly just under the same conditions 
as in \cite{KS}. The result
(Eq.(\ref{1.5})) agrees with Eq.(3.6) in \cite{BK2} which is input 
formula for analysis in \cite{BK2}.

\appendix

\section{Radiation in crystals}

\renewcommand{\theequation}{A.\arabic{equation}}

Separation of coherent and incoherent radiation in oriented single crystals
was considered using different approaches in
\cite{BG2},\cite{BG3}
and in \cite{BKS6}, \cite{BKS7}.
We give here a sketch of the analysis in one-chain
approximation neglecting correlations due to collision of projectile with
different chains. In this case
\begin{eqnarray}
\hspace{-7mm}&&\left\langle \sum_{a \neq b}
m_i(\textbf{r}_{\perp}-\textbf{x}_a) m_j(\textbf{r}_{\perp}-\textbf{x}_b)
e^{-iq_{\parallel}\left(z_a-z_b\right)} \right\rangle
\nonumber \\
\hspace{-7mm}&& = \left\langle m_i\right\rangle_{\perp}
\left\langle m_j \right\rangle_{\perp} e^{-q_{\parallel}^2u^2}\sum_{a \neq b}
e^{-iq_{\parallel}\left(z_a^{(0)}-z_b^{(0)}\right)},
\label{1.9}
\end{eqnarray}
where $u$ is the amplitude of the thermal vibrations, in averaging over
the thermal vibrations we used the distribution
\begin{equation}
w(z_a)= \frac{1}{\sqrt{2\pi}u}
\exp \left(-\frac{(z_a-z_a^{(0)})^2}{2u^2}\right).
\label{1.10a}
\end{equation}
The sum in the r.h.s. of (\ref{1.9})
one can present as
\begin{equation}
\sum_{a \neq b}
\exp(-iq_{\parallel}\left(z_a^{(0)}-z_b^{(0)})\right)
= - N_c + \left|\sum_a \exp(iq_{\parallel}z_a^{(0)})\right|^2,
\label{1.10}
\end{equation}
where
\begin{eqnarray}
\hspace{-7mm}&& \sum_a \exp(iq_{\parallel}z_{a}^{(0)})
=\sum_{n=-\infty}^{\infty}\exp(inq_{\parallel}d)
=2\pi\sum_{k=-\infty}^{\infty}\delta(q_{\parallel}d-2\pi k), 
\nonumber \\
\hspace{-7mm}&&
\left|\sum_a \exp(iq_{\parallel}z_{a}^{(0)}) \right|^2
=\frac{N_c}{d} 2\pi
\sum_{k=-\infty}^{\infty}\delta(q_{\parallel}-2\pi \frac{k}{d}),
\label{1.11}
\end{eqnarray}
where $d$ is the distance between atoms forming the chain.
Substituting Eqs.(\ref{1.9})-(\ref{1.11}) in Eq.(\ref{1.2}) we get
\begin{eqnarray}
\hspace{-7mm}&& \left\langle M_iM_j^{\ast} \right\rangle = 
N_c\left(\left\langle m_im_j \right\rangle_{\perp}
- \exp(-q_{\parallel}^2u^2)
\left\langle m_i \right\rangle_{\perp}\left\langle m_j \right\rangle_{\perp}\right)
\nonumber \\
\hspace{-7mm}&&+ N_c\frac{2\pi}{d}\exp(-q_{\parallel}^2u^2)
\left\langle m_i \right\rangle_{\perp}\left\langle m_j
\right\rangle_{\perp}
\sum_{k=-\infty}^{\infty}\delta(q_{\parallel}-2\pi \frac{k}{d}).
\label{1.12}
\end{eqnarray}
This expression agrees with Eq.(11) in \cite{BG2}. The incoherent term in
Eq.(\ref{1.12}) ($\propto N_c$)
coincide with the incoherent term in Eq.(\ref{1.7}) if $q_{\parallel}u \ll 1$.
This is true if the condition Eq.(\ref{1.3}) ($l_f/d \gg 1$) is satisfied.

\end{document}